\documentstyle[aps]{revtex}
\begin{document}
\title{ Factored coset models: A unifying approach to different bosonization schemes }
\author{A N Theron, F G Scholtz and H B Geyer}
\address{
Institute of Theoretical Physics,\\
University of Stellenbosch, 7600 Stellenbosch, South Africa}

\maketitle

\begin{abstract}
We discuss various bosonization schemes from a path integral perspective.
Our analysis shows that the existence of different bosonization schemes,
such as  abelian bosonization of non-abelian models and non-abelian
bosonization of fermions with colour and flavour indices, can be understood
as  different ways of factoring out a dynamically trivial coset which
contains the fermions. From this perspective follows the importance of the
coset model in ensuring the correct superselection rules on the bosonic
level.
\end{abstract}


\vspace{.7cm}


\bigskip



\noindent{\small {\bf Keyword abstract:} Path integral, Wess-Zumino-Witten
model, abelian bosonization, non-abelian bosonization, Thirring model\newline
}

\section{\protect\small Introduction}

{\small Bosonization has a long history in two dimensional quantum field
theory. The subject originated with the work of Coleman and Mandelstam \cite
{1} on abelian models in the middle seventies (see also Ref. \cite{lp}), and
was soon applied to the non-abelian case \cite{2,3}. This generalization, 
referred to hereafter as ``earlier non-abelian bosonization", however,
obscures the non-abelian symmetries of the original fermionic model. Since
Witten's introduction of a non-abelian bosonization scheme which maintains
in a manifest way the fermionic symmetries on the bosonic level \cite{4},
the ``earlier non-abelian" bosonization has remained somewhat in the
background. }

{\small The bosonization scheme of Witten is based on the equivalence of
free fermions to the Wess-Zumino-Witten (WZW) \cite{4} model and has been
applied with great success to a number of field theoretical problems, in
particular to massive fermions carrying both flavour and colour indices \cite
{5,6}. Such fermions may be bosonized in terms of either a $U(N_f N_c)$ WZW
model or in terms of a $SU(N_f) \times SU(N_c) \times U(1)$ WZW model. A
subtlety arises here since, although the last option is desirable because it
separates the colour and flavour degrees of freedom, it has been pointed out
in Refs. \cite{5,6}, that the resulting bosonic image of the fermionic mass
term is incorrect. }

{\small Bosonization has been extensively discussed from the path integral
point of view. An approach has been pioneered in Ref. \cite{7} where
interacting fermionic models are bosonized by ``decoupling" the fermions via
a chiral transformation. More recently, interest in deriving the
bosonization rules or dictionary with path integrals has been revived \cite
{8,9,10}. In Ref. \cite{8} it was shown that bosonization can be viewed as a
type of duality transformation, and in Ref. \cite{9} it was derived by first
introducing a gauge symmetry, followed by choosing an appropriate gauge. In
Ref. \cite{10} it was shown that the bosonization rules can be derived by
factoring the fermions out as part of a trivial coset model. In this manner
the complete bosonization dictionary for both abelian and non-abelian
bosonization was derived. A particular advantage of this approach is that
the superselection rules are embodied in the dynamically trivial coset
model. }

{\small Despite extensive work on bosonization some gaps still exist in the
literature. ``Earlier non-abelian bosonization", the equivalence of the
massless $SU(2)$ Thirring model to the sine-Gordon and free boson model (not
to be confused with the equivalence of the massive Thirring model and the
sine-Gordon model \cite{1,naon}), and the different non-abelian bosonization
schemes for coloured and flavoured fermions have not been derived with path
integrals. The above mentioned subtlety in the $SU(N_f) \times SU(N_c)
\times U(1)$ non-abelian scheme for mass bilinears has not been resolved at
all. In this paper we present a path integral derivation of the ``earlier
non-abelian bosonization" and show the equivalence of the $SU(2)$ Thirring
model with the sine-Gordon model. We also discuss non-abelian bosonization
for fermions with colour and flavour indices and show how to derive both the 
$U(N_f N_c)$ and $SU(N_f)\times SU( N_c)\times U(1)$ schemes with path
integrals. Our method also leads to the correct form for the boson image of
the fermionic mass terms in the $SU(N_f)\times SU( N_c)\times U(1)$ scheme. }

{\small These results are obtained using the formalism of Ref. \cite{10},
where the fermions are factored out as part of a coset model which is
dynamically trivial. Indeed, we show that the above mentioned bosonization
schemes correspond to factoring out the fermions as part of different coset
models. The formalism of Ref. \cite{10} therefore provides us not only with
a convenient way of deriving the bosonization dictionary but also with a way
of understanding the relation between different bosonization schemes and
thus act as a unifying framework for the different schemes. Furthermore the
coset model explicitly encapsulates the superselection rules of the
different schemes. On the canonical level the coset model that is factored
out corresponds to either a finite dimensional or a discrete Hilbert space
that must be retained to ensure the correct superselection rules. We show
that this is of particular importance for the ``earlier non-abelian"
bosonization and for the non-abelian $SU(N_f)\times SU( N_c)\times U(1)$
scheme where neglect of this Hilbert space actually results in the wrong
boson image for the fermionic mass terms. }

{\small The outline of this paper is a follows. In Section 2 we briefly
review the formalism of Ref. \cite{10} for the sake of completeness. In
Section\ 3 we discuss the ``earlier non-abelian bosonization'' from a path
integral perspective and in Section\ 4 apply this to the $SU(2)$ Thirring
model to re-derive the equivalence of this model to the sine-Gordon model.
In so doing we give a path integral derivation of this interesting result.
In Section\ 5 we discuss non-abelian bosonization, followed by some
conclusions. }

\section{\protect\small Bosonization by factoring out the fermions}

{\small We briefly review the main results of Ref. \cite{10}. Working in
two-dimensional Minkowski space we start with the generating functional of
free Dirac fermions with source terms (or external fields) for the currents
and mass terms. }

{\small 
\begin{equation}  \label{1}
Z[A_{+},\eta, {\bar \eta} ] = \int \! D \bar{\psi} D \psi \exp\{i\int \! d^2
x \,[\bar{\psi} i \gamma^{\mu} \partial_{\mu} \psi + \psi^{\dagger}_-
A_{+}\psi_- + \eta \psi^{\dagger}_{+}\psi_{-} + {\ \bar \eta}
\psi^{\dagger}_{-} \psi_{+}] \} \quad.
\end{equation}
}

{\small Light cone components are used and for simplicity we first consider
an $A_+$ dependence only, returning to the more general case later. The
first step is the introduction of bosonic degrees of freedom in the path
integral. This is accomplished by inserting the following identity in the
path integral 
\begin{eqnarray}  \label{2}
1&=&\int \! D B_{+} \delta(B_{+}) \exp\{i \int \! d^2 x\, \psi_-^{\dagger}
B_{+}\psi_- \}  \nonumber \\
& =&\int \! D B_{+} D \lambda_{-} \exp\{i\int \! d^2 x\,[ \psi_-^{\dagger}
B_{+}\psi_- + B_{+} \lambda_{-}] \}\quad .
\end{eqnarray}
Inserting this identity in eq. (\ref{1}) and shifting $B_+ \rightarrow B_+ +
A_+$, we have }

{\small 
}

{\small 
\begin{equation}  \label{3}
Z[A_{+},{\bar \eta},\eta] = \int \! D \bar{\psi} D \psi D B_{+} D
\lambda_{-} e^{i\int \! d^2 x\, L }
\end{equation}
with 
\begin{equation}
L= \psi_+^{\dagger} i \partial_- \psi_+ + \psi_-^{\dagger} i \partial_+
\psi_- + \psi_-^{\dagger} B_{+} \psi_- +B_+ \lambda_{-} - A_{+} \lambda_{-}
+ \eta \psi^{\dagger}_{+}\psi_{-} + {\ \bar \eta} \psi^{\dagger}_{-}
\psi_{+} \quad.
\end{equation}
}

{\small The aim is to factor the fermionic part of the generating functional
out as part of a constrained fermionic model which is topological and thus
contains no dynamics. This is achieved by the following change of variables
in the path integral }

{\small 
\begin{equation}  \label{4}
\begin{array}{rclcrcl}
\psi_{-} & \rightarrow & e^{-i \theta (x) } \psi_{-} & \quad \quad & \psi_{+}
& \rightarrow & \psi_{+} \nonumber \\ 
\psi^{ \dagger }_{-} & \rightarrow & \psi^{\dag}_{-} e^{ i \theta (x) } & 
\quad\quad & \psi^{ \dagger}_{+} & \rightarrow & \psi^{\dag}_{+} \quad,
\end{array}
\end{equation}
\begin{equation}  \label{5}
\lambda_{-} = \alpha \partial_{-} \theta ,
\end{equation}
where $\alpha$ is a number to be specified later. }

{\small The change of variables (\ref{4}) is a chiral transformation and the
fermionic measure is not invariant under this transformation. This results
from the chiral anomaly which appears in the path integral formalism as a
non-invariance of the fermionic measure, as first shown by Fujikawa \cite{11}
for an infinitesimal transformation. For the finite rotation required in
expression (\ref{4}), the Jacobian reads \cite{7,10} }

{\small 
\begin{equation}  \label{6}
J_{{\rm {F}}} = \exp \Bigl(\frac{i}{ \pi}\int \!d^2x( \frac{1}{2}
\partial_{-}\theta \partial_{+} \theta + \partial_{-}\theta B_{+})\Bigr)\; .
\end{equation}
}

{\small The determinant for the change of variables (\ref{5}) is $%
\det(\alpha \partial_-)$ and it may be written as a functional integral over
Grassmann variables (ghosts) }

{\small 
\begin{equation}  \label{7}
\int \! D {b}_+ D c_+ e^{i\int \! d^2 x \, {b}_+ \partial_{-} c_+} \quad.
\end{equation}
}

{\small This determinant is actually irrelevant for evaluating the
generating functional (\ref{1}) as it only affects the normalization. It is,
however, important if we are interested in correlation functions of the
energy momentum tensor and the central charge of the model. }

{\small The generating functional (\ref{3}) now reads 
\begin{equation}  \label{8}
Z[A_+,{\bar \eta},\eta] = \int \! D \bar{\psi} D \psi D B_{+} D \theta D b_+
D c_+ e^{i\int \! d^2 x {\ L}_{{\rm cf}} + {\ L}_{{\rm B}} + {\ L}_{{\rm %
source}} } \, ,  \nonumber
\end{equation}
where 
\begin{eqnarray}  \label{11}
{\ L}_{{\rm cf}} &=& \psi_+^{\dagger} i \partial_- \psi_+ + \psi_-^{\dagger}
i \partial_+ \psi_- + \psi_-^{\dagger}( B_{+} - \partial_+\theta) \psi_- + {b%
}_+ \partial_{-} c_+ \, , \\
{\ L}_{{\rm B}} &=& \frac{1}{2\pi}\partial_+\theta\partial_-\theta +\frac{1}{
\pi}B_+\partial_-\theta + \alpha B_+ \partial_-\theta \, , \\
{\ L}_{{\rm source}} &=& -\alpha A_{+} \partial_{-} \theta + \eta
\psi^{\dagger}_{+} e^{i \theta}\psi_{-} + {\ \bar \eta} \psi^{%
\dagger}_{-}e^{-i\theta} \psi_{+} \quad.  \label{12}
\end{eqnarray}
$L_{{\rm B}}$ originates from the fermionic Jacobian (\ref{6}) and the $%
\lambda_- B_+$ term of eq. (\ref{3}). We note that if we choose $\alpha = 
\frac{-1}{\pi}$ the $B_+$ dependence dissapears completely from ${L}_{{\rm B}%
}$ and the generating functional (\ref{8}) factorises into that of a
constrained fermionic model and a free massless boson model. (The $\theta$
dependence in the fermionic part dissapears under the shift $B_+ \rightarrow
B_+ - \partial_+ \theta$.) The fermionic and bosonic Lagrangians are still
coupled through the sources, but upon functional differentiation and setting
the sources ($\eta$ and $\bar{\eta}$) equal to zero we achieve complete
decoupling. To summarize, the generating functional factorizes as 
}

{\small 
\begin{equation}  \label{13}
Z[A_{+},{\bar{\eta}},\eta ]_{\bar{\eta},\eta =0} = Z_{{\rm cf}} Z_{{\rm B}}
[A_{+},{\bar{\eta}},\eta ]_{\bar{\eta},\eta =0}
\end{equation}
where 
\begin{equation}  \label{14}
Z_{{\rm cf}} = \int \! D \bar{\psi} D \psi D B_{+} D b_+ D c_+ e^{iS_{{\rm cf%
}}}
\end{equation}
with 
\begin{equation}  \label{lc}
S_{{\rm cf}}=\int \! d^2 x \, [ \psi_+^{\dagger} i \partial_- \psi_+ +
\psi_-^{\dagger} i \partial_+ \psi_- + \psi_-^{\dagger} B_{+} \psi_- + {b}_+
\partial_{-} c_+]
\end{equation}
and 
\begin{equation}
Z_{{\rm B}}[A_{+},{\bar{\eta}},\eta ] = \int \! D \theta e^{iS_{{\rm B}}}
\end{equation}
with 
\begin{equation}
S_{{\rm B}}= \int \! d^2 x \,[ \frac{1}{2 \pi} \partial_{-} \theta
\partial_{+} \theta + \frac{1}{\pi} A_{+} \partial_{-}\theta + \eta
\psi^{\dagger}_{+} e^{i \theta}\psi_{-} + {\ \bar \eta} \psi^{%
\dagger}_{-}e^{-i\theta} \psi_{+} ] \quad.  \label{15}
\end{equation}
}

{\small 
}

{\small Eqs. (\ref{13}) to (\ref{15}) are the desired results.  If we wish
to calculate current-current correlation functions in the generating
functional (\ref{3}), we may calculate these correlators instead with the
generating functional (\ref{15}).  The fermions are decoupled as part of a
constrained fermionic model which contains no dynamics. It is a topological
model that only contains information about the superselection rules of the
original fermions. }

{\small To calculate correlators of mass bilinears in (\ref{3}) we have to
differentiate with respect to $\eta$ and $\bar{\eta}$ and put these sources
equal to zero. Using the result (\ref{8}) we obtain for example }

{\small 
\begin{eqnarray}  \label{16}
\frac{\delta^2 Z[A_+=0,{\bar\eta},\eta]}{\delta {\bar \eta}(x_1) \;\delta {\
\eta}(x_2)}|_{{\bar \eta}= \eta =0} &=& \langle\;
\psi^{\dagger}_{-}\psi_{+}(x_1) \psi^{\dagger}_{+} \psi_{-}(x_2)\;\rangle_{%
{\rm {free\,\, fermion}}}  \nonumber \\
& =& \langle\; \psi^{\dagger}_{-}\psi_{+}(x_1) \psi^{\dagger}_{+}
\psi_{-}(x_2)\;\rangle_{{\rm {coset}}} \langle\;e^{- i \theta(x_1)}e^{ i
\theta(x_2)}\;\rangle_{{\rm {free\,\, boson}}}\,\, ,
\end{eqnarray}
where $\langle\,\, \rangle_{{\rm coset}} $ denotes the correlation function
of mass bilinears evaluated in the coset model. Because of the trivial
nature of the coset this quantity is a constant, so that we obtain the
result that the free fermion correlator is proportional to the correlators
of exponentials of free boson fields. In this way one obtains the standard
bosonization rule that fermionic mass terms may be represented by the
exponentials of free bosonic fields. }

{\small The fermionic coset model \cite{12} is a model with a discrete
Hilbert space. It is a conformally invariant model with central charge $c=0$ 
\cite{12}, and therefore has only $h=0$ primary fields. This is another way
of noting that the correlators of primary fields are constant. }

{\small We would like to stress that eq. (\ref{14}) should be interpreted as
a gauge fixed fermionic coset model in the light cone gauge, that is, both
left and right currents are constrained to zero. The light cone gauge was
induced by the choice of the identity (\ref{2}). Indeed, an alternative
choice of the identity (\ref{2}) leads to the Lorentz gauge, as explained
below. Alternatively, as explained in Ref. \cite{10}, a coset model that is
not gauge fixed may be obtained by inserting the quantity %
\begin{eqnarray}  \label{a1}
&& \int \! D B_{\mu} \delta( \epsilon^{\mu \nu} \partial_{\nu} B_{\mu} )
\exp\{i \int \! d^2x \,{\bar \psi}\gamma^{\mu} B_{\mu} \psi \}  \nonumber \\
&=& \int \! D B_{\mu} D \lambda \exp\{i \int \! d^2x \,[{\bar \psi}%
\gamma^{\mu} B_{\mu} \psi + \lambda\epsilon^{\mu \nu} \partial_{\nu} B_{\mu}
] \} \quad
\end{eqnarray}
into the generating functional 
\begin{equation}  \label{a2}
Z[A_{\mu} ] = \int \! D \bar{\psi} D \psi \exp\{i\int \! d^2 x\, \bar{\psi}
i \gamma^{\mu}( \partial_{\mu} + A_{\mu})\psi \} \quad.
\end{equation}
}

{\small This quantity may be inserted into the path integral without
altering the physics \cite{10}. It amounts to gauging the fermions and
constraining the connection $B_{\mu}$ to be flat \cite{8}. }

{\small The fermions are now factored out by the $\gamma_5$ transformation }

{\small 

\begin{eqnarray}  \label{a3}
\psi & \rightarrow & e^{-i \gamma_5 \theta (x) } \psi\,,  \nonumber \\
{\bar \psi} & \rightarrow & {\bar \psi} e^{-i \gamma_5 \theta(x) } \,, 
\end{eqnarray}

\begin{eqnarray}  \label{a4}
\lambda(x) & \rightarrow & \frac{1}{\pi} \theta(x)\; .
\end{eqnarray}
The Jacobian associated with the change of variables (\ref{a3}) reads 
\begin{equation}  \label{a5}
J_{{\rm {F}}} = \exp\{\frac{i}{ \pi}\int \!d^2x( \frac{1}{2}
\partial_{\mu}\theta \partial^{\mu} \theta + \epsilon^{\mu \nu}
\partial_{\nu }\theta ( B_{\mu} + A_{\mu}))\}\; ,
\end{equation}
while we ignore the Jacobian associated with (\ref{a4}) which is just a
divergent constant. We obtain }

{\small 
\begin{equation}  \label{a6}
Z[A_{\mu} ] = \int \! D \bar{\psi} D \psi D B_{\mu} D \theta\, e^{i\int \!
d^2 x L}
\end{equation}
with 
\begin{equation}
L= \bar{\psi} i \gamma^{\mu} \partial_{\mu} \psi +{\bar \psi}\gamma^{\mu} (
A_{\mu} + B_{\mu} +\epsilon_{\mu \nu} \partial^{\nu}\theta)\psi +\frac{1}{2
\pi} \partial_{\nu}\theta \partial^{\nu}\theta +\frac{1}{\pi}%
A_{\mu}\epsilon^{\mu \nu} \partial_{\nu}\theta \; .
\end{equation}
Shifting the $B_{\mu}$ field to $B_{\mu} \rightarrow B_{\mu} + A_{\mu} +
\epsilon_{\mu \nu} \partial_{\nu} \theta $ yields the desired result -- a
coset model which has not been gauge fixed and a free bosonic field. }

{\small To obtain a coset model in the Lorentz gauge, one replaces the
identity (\ref{2}) by %
\begin{eqnarray}  \label{a7}
1&=&\int \! D B_{\mu} \delta(B_{\mu}) \exp\{i \int \! d^2 x\, {\bar \psi}%
\gamma^{\mu} B_{\mu}\psi \}  \nonumber \\
& =&\int \! D B_{\mu} D \lambda_{\mu} \exp\{i\int \! d^2 x\,[ {\bar\psi}
\gamma^{\mu} B_{\mu}\psi + B_{\mu} \lambda^{\mu}] \}\quad,
\end{eqnarray}
which is manifestly covariant, and insert it into the generating functional (%
\ref{a2}). We again perform the transformation (\ref{a3}) but replace
transformation (\ref{a4}) by %
%
\begin{equation}  \label{a8}
\lambda_{\mu}= \partial_{\mu} \eta + \frac{1}{\pi}\epsilon_{\mu \nu}
\partial^{\nu}\theta
\end{equation}
with Jacobian $\det (\frac{1}{ \pi } \partial_{\mu}\partial^{\mu})$. This
yields the result }

{\small 
\begin{equation}  \label{a9}
Z[A_{\mu} ] = \int \! D \bar{\psi} D \psi D B_{\mu} D\eta D \theta
\det(\partial_{\mu}\partial^{\mu}) e^{i\int \! d^2 x\, L}
\end{equation}
with 
\begin{equation}
L= \bar{\psi} i \gamma^{\mu} \partial_{\mu}\psi + {\bar \psi}\gamma^{\mu}
(B_{\mu} + A_{\mu } + \epsilon_{\mu \nu}\partial^{\nu} \theta ) \psi
+\partial^{\mu}\eta B_{\mu} +\frac{1}{2 \pi}\partial^{\mu}\theta
\partial_{\mu}\theta + \frac{1}{\pi} \epsilon^{\mu
\nu}A_{\mu}\partial_{\nu}\theta \quad.
\end{equation}
}

{\small Shifting the $B_{\mu}$ field to $B_{\mu} \rightarrow B_{\mu} +
\epsilon_{\mu \nu} \partial^{\nu}  \theta $,  and performing a partial
integration it is realized that $\eta$ is just the Lagrange multiplier that
enforces the Lorentz gauge condition $\partial^{\mu} B_{\mu}$. The
determinant $\det (\frac{1}{ \pi} \partial_{\mu}\partial^{\mu})$ is the
Fadeev-Popov determinant associated with the Lorentz gauge and may be
written as an integral with ghost fields similar to eq. (\ref{7}). The
appearance of $A_{\mu}$ in the coset sector is spurious; functional
differentiation at $A_{\mu}=0$ yields no contribution from this sector, 
because correlators of coset currents vanish. }

{\small This concludes our review of abelian bosonization in the factored
coset approach. The trivial model may be factored out in the light cone
gauge  (\ref{lc}), the Lorentz gauge (\ref{a9}), or in a form (\ref{a6})
that has not been gauge fixed at all. }

\section{\protect\small Earlier non-abelian bosonization}

{\small We now consider the non-abelian case and apply the formalism of the
previous section. For simplicity we only consider a model with fermions in
the fundamental representation of $SU(2)$, %
\begin{equation}  \label{17}
Z[A_{+}^a ] = \int \! D \bar{\psi} D \psi \exp\{i \int \! d^2 x \,[\bar{\psi}
i \gamma^{\mu} \partial_{\mu} \psi + \psi^{\dagger}_- A_{+}^a t^a \psi_- ]
\} \; ,
\end{equation}
%
%
with $t^a=\sigma^a/2$ and $\sigma^a$ the Pauli matrices. We call this
additional index a flavour index.  We constrain each flavour of fermion
separately into a $U(1)$ coset model by introducing the identity %
\begin{equation}  \label{18}
1=\int \! D B_{+}^1 \delta(B_{+}^1)D B_{+}^2 \delta(B_{+}^2) \exp\{i \int \!
d^2 x\,[ \psi_-^{\dagger 1} B_{+}^1\psi_-^1 + \psi_-^{\dagger 2} B_{+}^2
\psi_-^2] \} \quad ,
\end{equation}
where we have written the flavour index explicitly. We introduce two
Lagrange multipliers $\lambda^i_+$ to lift the Dirac deltas into the action,
perform on each flavour a $U(1)$ chiral transformation 
\begin{equation}  \label{19}
\begin{array}{rclcrcl}
\psi_{-}^a & \rightarrow & e^{-i \theta^a (x) } \psi_{-}^a & \quad \quad & 
\psi_{+}^a & \rightarrow & \psi_{+}^a \nonumber \\ 
\psi^{ \dagger a}_{-} & \rightarrow & \psi^{\dagger a}_{-} e^{ i \theta^a
(x) } & \quad\quad & \psi^{ \dagger a}_{+} & \rightarrow & \psi^{\dagger
a}_{+} \quad,
\end{array}
\end{equation}
%
and then transform the Lagrange multipliers to %
\begin{equation}  \label{20}
\lambda^a=\frac{1}{ \pi} \partial_-\theta^a\quad.
\end{equation}
}

{\small Each flavour has an associated anomalous Jacobian (\ref{6}), so that
we obtain the result (after lifting the Jacobians associated with (\ref{20})
with ghosts and making the appropriate shifts on the $B$ fields) %
%
%
\begin{eqnarray}  \label{21}
Z[A_{\mu} ] & = & \int \! D \bar{\psi} D \psi DB_{+}^1 DB_{+}^{2} D \theta^1
D \theta^2 D b_+^1 Dc_+^1Db_+^2Dc_+^2  \nonumber \\
&&e^{i\int \!d^2x[L_{{\rm cf}} +L_{{\rm B}} +L_{{\rm source}}] } \quad.
\end{eqnarray}
Here the various Lagrangians are defined as 
\begin{equation}  \label{22}
L_{{\rm cf}}= \sum_{j=1,2} [\bar{\psi^j} i \gamma^{\mu} \partial_{\mu}
\psi^j + B_{\mu}^j{\bar \psi}^j \gamma^{\mu} \psi^j +b_+^j\partial_-c_+^j ]
\, ,
\end{equation}
\begin{equation}  \label{23}
L_{{\rm B}}= \sum_{j=1,2} \frac{1}{2 \pi} \partial_{\nu}\theta^j
\partial^{\nu}\theta^j
\end{equation}
\,, 
\begin{eqnarray}  \label{24}
L_{{\rm source}}&=& \frac{1}{2\pi}A_{+}^3( \partial_{-}\theta_1 -
\partial_{-}\theta_2)  \nonumber \\
&& +\frac{1}{2}( \psi^{\dagger 1}_- \psi^2_- e^{-i \theta^1 + i \theta^2} +
\psi^{\dagger 2}_- \psi^1_- e^{i \theta^1 - i \theta^2}) A_{+}^{(1)} 
\nonumber \\
&+&\frac{i}{2}( \psi^{\dagger 2}_- \psi^1_- e^{i \theta^1 - i \theta^2}
-\psi^{\dagger 1}_- \psi^2_- e^{-i \theta^1 + i \theta^2}) A_{+}^{(2)} \quad.
\end{eqnarray}
%
The constrained model ({\ref{22}) is dynamically trivial. The central charge
of this model is zero and correlators of the fermion fields are constant. }}

{\small The first two terms of ${L}_{{\rm source }}$ originate from the
Jacobian (\ref{6}) because $A_+^3$ couples to the $u(1)$ currents. Since
current correlators are obtained by differentiating with respect to the
sources and setting them equal to zero, one notes that the bosonic images of
the fermionic currents may be read off immediately from ${L}_{{\rm source }}$%
. These rules coincide with the ``old non-abelian bosonization rules" of
Banks et al. \cite{2}, thus yielding a first path integral derivation. Of
importance is the appearance of the coset fields $\psi^i$ in these rules. As
mentioned, the correlators of these fields are constant -- on the second
quantized level these fields are the path integral analogue of fermionic
operators without a space-time dependence. }

{\small In writing down the bosonic image of the fermionic currents, the
authors of Ref. \cite{2} need to  introduce fermionic operators with no
space-time dependence, denoted by $\chi$ in Ref. \cite{2}. In our approach
the coset fields play the role of these constant fermion operators. The
constrained fermion approach to bosonization therefore provides a natural
framework to understand the need to introduce these constant fermionic
operators in the bosonization dictionary. }

{\small The boson images of the mass bilinears are likewise obtained by
introducing the appropriate source terms in (\ref{24}). }

\section{\protect\small Equivalence of the $SU(2)$ Thirring and sine-Gordon
models}

{\small In this section we apply the preceding formalism to the $SU(2)$
massless Thirring model. It was shown in Ref. \cite{2} from ``earlier
non-abelian bosonization" that this model is equivalent to a free massless
boson and a sine-Gordon model. We discuss here this important result from a
path integral point of view. }

{\small The Lagrangian of the massless $SU(2)$ Thirring model is given by %
%
\begin{equation}  \label{3.1}
L_{{\rm T}}= {\ \bar \psi}i\gamma^{\mu}\partial_{\mu} \psi - g^2
j_{\mu}^aj^{\mu a }\quad,
\end{equation}
where $\psi$ is in the fundamental representation of $SU(2)$ and $j_{\mu}^a={%
\ \bar \psi} \gamma_{\mu}t^a \psi$ with $t^a$ given below eq. (\ref{17}). }

{\small The four-point interaction is eliminated by introducing an auxiliary
field $A_{\mu}^a$ to yield equivalently %
%
\begin{equation}  \label{3.3}
L_{{\rm T}}= {\ \bar \psi}i\gamma^{\mu}\partial_{\mu} \psi + 2g j_{\mu}^a
A^{\mu a } +A_{\mu }^a A^{\mu a}\quad.
\end{equation}
%
%
%
}

{\small Since both left and right components of the currents are present we
find it convenient to use the formalism based on inserting the quantity (\ref
{a1}) for both flavours into the generating functional. The Lagrangian then
reads (we write the flavour indices explicitly) 
\begin{eqnarray}  \label{3.6}
L_{{\rm T}} &=& {\ \bar \psi}i\gamma^{\mu}\partial_{\mu} \psi +{\bar \psi}%
^1\gamma^{\mu}(B_{\mu}^1 + g A_{\mu}^0)\psi^1  \nonumber \\
&+&{\bar \psi}^2\gamma^{\mu}(B_{\mu}^2 - g A_{\mu}^0)\psi^2 +g {\bar \psi}%
^2\gamma^{\mu} A_{\mu}^{(-)}\psi^1 +g {\bar \psi}^1\gamma^{\mu}
A_{\mu}^{(+)}\psi^2  \nonumber \\
&+& A_{\mu}^{(0)} A^{\mu (0)} +A_{\mu}^{(+)} A^{\mu (-)} +\frac{1}{\pi}%
\theta^1\epsilon^{\mu \nu}\partial_{\nu}B_{\mu}^1 +\frac{1}{\pi}%
\theta^2\epsilon^{\mu \nu}\partial_{\nu}B_{\mu}^2)
\end{eqnarray}
where $A^{(\pm)}_{\mu}=A^{(1)}_{\mu} \mp i A^{(2)}_{\mu}$. The $B_{\mu}^i$
and $\theta^i$ fields carry a flavour index as in the previous section. The
auxilliary field $A_{\mu}$ plays a similar role as the source $A_{\mu}$ of
the previous section. }

{\small The fermions are constrained by the chiral rotation 
\begin{equation}  \label{3.7}
\psi^j \rightarrow e^{-i \gamma^5 \theta^j}\psi^j \quad ,\quad {\bar \psi}^j
\rightarrow {\bar \psi}^j e^{-i \gamma^5 \theta^j} \, ,
\end{equation}
to obtain, (taking again the Jacobian (\ref{a5}) into account) %
%
\begin{eqnarray}  \label{3.8}
L_{{\rm T}} &=& {\ \bar \psi}i\gamma^{\mu}\partial_{\mu} \psi +{\bar \psi}%
^1\gamma^{\mu}(B_{\mu}^1 + g A_{\mu}^0 +\epsilon_{\mu
\nu}\partial^{\nu}\theta^1) \psi^1  \nonumber \\
&+&{\bar \psi}^2\gamma^{\mu}(B_{\mu}^2 - g A_{\mu}^0 +\epsilon_{\mu
\nu}\partial^{\nu}\theta^2) \psi^2 +g {\bar \psi}^2 e^{i \gamma^5 \theta^2}
\gamma^{\mu}e^{i \gamma^5 \theta^1} \psi^1 A_{\mu}^{(-)}  \nonumber \\
&+&g {\bar \psi}^1 e^{i \gamma^5 \theta^1} \gamma^{\mu} e^{i \gamma^5
\theta^2} \psi^2 A_{\mu}^{(+)} +A_{\mu}^{(0)} A^{\mu (0)} +A_{\mu}^{(+)}
A^{\mu (-)}  \nonumber \\
&+&\frac{1}{2 \pi}\partial_{\mu}\theta^1\partial^{\mu}\theta^1 +\frac{1}{2
\pi}\partial_{\mu}\theta^2 \partial^{\mu}\theta^2 +\frac{g}{\pi}%
\epsilon^{\mu \nu}\partial_{\nu}\theta^1 A_{\mu}^{(0)}  \nonumber \\
&-&\frac{g}{\pi}\epsilon^{\mu \nu}\partial_{\nu}\theta^2 A_{\mu}^{(0)} \quad.
\end{eqnarray}
}

{\small Shifting, as before, the $B_{\mu}^i$ field and performing the
integration over the auxilliary field $A_{\mu}$, we obtain 
\begin{eqnarray}  \label{3.9}
L_{{\rm T}} &=& L_{{\rm coset}} + +g^2 {\bar \psi}^2 e^{i \gamma^5 \theta^2}
\gamma^{\mu}e^{i \gamma^5 \theta^1} \psi^1 {\bar \psi}^1 e^{i \gamma^5
\theta^1} \gamma^{\mu} e^{i \gamma^5 \theta^2} \psi^2  \nonumber \\
&+&\left(\frac{1}{2 \pi} + \frac{g^2}{4\pi^2}\right)
\partial_{\mu}\theta^1\partial^{\mu}\theta^1 +\left(\frac{1}{2 \pi} + \frac{%
g^2}{4\pi^2}\right) \partial_{\mu}\theta^2 \partial^{\mu}\theta^2  \nonumber
\\
&-&\frac{g^2}{2\pi^2} \partial_{\mu}\theta^1\partial^{\mu}\theta^2 \quad,
\end{eqnarray}
where 
\begin{equation}  \label{3.9a}
L_{{\rm coset}}= {\ \bar \psi}i\gamma^{\mu}\partial_{\mu} \psi +{\bar \psi}%
^1\gamma^{\mu}B_{\mu}^1 \psi^1 +{\bar \psi}^2\gamma^{\mu}B_{\mu}^2 \psi^2
\quad.
\end{equation}
This may be simplified somewhat by the change of variables 
%
\begin{equation}  \label{3.10}
\eta=\theta^1+\theta^2 \quad \rho=\theta^1-\theta^2 \quad,
\end{equation}
to obtain %
%
\begin{eqnarray}  \label{3.11}
L_{{\rm T}} &=& L_{{\rm coset}} + g^2 {\bar \psi}^2 e^{-i \gamma^5 \rho}
\gamma^{\mu} \psi^1 {\bar \psi}^1 e^{i \gamma^5 \rho} \gamma^{\mu} \psi^2 
\nonumber \\
&+&\frac{1}{4 \pi} \partial_{\mu}\eta\partial^{\mu}\eta +\left(\frac{1}{4 \pi%
} + \frac{g^2 }{4\pi^2}\right)\partial_{\mu}\rho\partial^{\mu}\rho\quad.
\end{eqnarray}
$L_{{\rm coset}}$ is defined in (\ref{3.9a}). The $\eta$ field corresponding
to the $U(1)$ symmetry of the Thirring model (it may be readily verified
that $\epsilon^{\mu \nu} \partial_{\mu}\eta$ is the bosonic image of the $%
U(1)$ current ${\bar \psi}^1\gamma^{\mu}\psi^1 + {\bar \psi}%
^2\gamma^{\mu}\psi^2 $ by adding the relevant source term to the action)
completely decouples from the rest. }

{\small The only way of simplifying the generating functional further is by
making a perturbative expansion of the interaction term of (\ref{3.11}) in a
manner similar as in Refs. \cite{naon,mnt}. The interaction term may be
written in spinor components as 
\begin{equation}  \label{3.12}
L_{{\rm int}}= 2g^2 {\psi}^{\dagger 2}_- \psi^1_- \psi^{\dagger 1 }_+ \psi^2
_+ e^{-2 i \rho} + 2 g^2 {\psi}^{\dagger 2}_+ \psi^1_+ \psi^{\dagger 1 }_-
\psi^2 _- e^{2 i \rho} \quad.
\end{equation}
(We ignore terms independent of $\rho$ as they contribute only a constant.)
Making an expansion of the generating functional in terms of $L_{{\rm int}}$
gives 
\begin{eqnarray}  \label{3.13}
Z_{{\rm T}}&=& \int \! D {\bar \psi}D \psi D B_{\mu}^1 D B_{\mu}^2 D \rho D
\eta e^{iS_{{\rm coset}} + i S_{{\rm free \,boson}} }  \nonumber \\
&&\,\,\times (1+ \int\!d^2 x \; 2g^2 ( \psi^{\dagger 2}_- \psi^1_-
\psi^{\dagger 1 }_+ \psi^2 _+ e^{-2 i \rho} + \psi^{\dagger 2}_+ \psi^1_+
\psi^{\dagger 1 }_- \psi^2 _- e^{2 i \rho}) +\ldots)\,\,,
\end{eqnarray}
where 
\begin{equation}
S_{{\rm free \,boson}} = \int \!d^2x[ \frac{1}{4 \pi} \partial_{\mu}\eta%
\partial^{\mu}\eta +(\frac{1}{4 \pi} + \frac{g^2 }{4\pi^2}%
)\partial_{\mu}\rho\partial^{\mu}\rho ] \quad.
\end{equation}
}

{\small The fermionic correlators that appear in the perturbative expansion (%
\ref{3.13}) yield a constant, because theys are evaluated in the coset
model. Using chiral selection rules we note that  the series (\ref{3.13}) is
just the  expansion of an exponent with argument ${\rm constant} \times \cos
(2 \rho)$ in the free bosonic model -- see also Refs.\cite{1,naon,mnt}. This
establishes the equivalence between the $SU(2)$ Thirring model and the
sine-Gordon model together with a free boson field (the $\eta$ field). }

\section{\protect\small Non-abelian bosonization}

{\small We now turn our attention to the non-abelian bosonization of Witten 
\cite{4}, where the free fermions (with a $U(N)$ chiral symmetry) are
bosonized in terms of a WZW model which realizes the non-abelian symmetry in
a manifest way. In ref. \cite{10} it was shown that this non-abelian
dictionary may be obtained by factoring out a $U(N)$ coset model instead of $%
N$ $U(1)$ coset models as discussed in the previous section. Apart from this
difference the procedure is similar. We briefly review the main results of 
\cite{10} concerning non-abelian bosonization. }

{\small We start again with the generating functional for the currents and
fermionic mass terms. The mass terms are chosen to be invariant under $U(N)$
gauge transformation, this implies that it is general enough to consider a
generating functional that only depends on one component $A_+$ for the
current because the $A_-$ may be gauge transformed to zero as explained in 
\cite{10}. }

{\small 
\begin{eqnarray}  \label{25}
Z[A_{+},{\bar \eta},\eta ]& = & \int \! D {\bar \psi} D \psi \exp\{i\int \!
d^2 x\,[ \psi_{-}^{\dagger} i \partial_{+} \psi_{-} + \psi_{+}^{\dagger} i
\partial_{-} \psi_{+}  \nonumber \\
& -&\psi_{-}^{\dagger} A_{+} \psi_{-} + \eta \psi_{+}^{\dagger} \psi_{-} + {%
\bar \eta} \psi_{-}^{\dagger} \psi_{+} ] \} \quad,
\end{eqnarray}
where $A_+=A_+^at^a$ with $t^a$ the generators of $SU(N)$ and $\psi$ in the
fundamental representation of $SU(N)$. We factor out a $U(N)$ coset model by
introducing the following identity %
%
\begin{eqnarray}  \label{26}
1 &=& \int \! D B_{+}^a \delta( B_{+}^a) DC_+\delta(C_+) \exp \{i\int \! d^2
x \, [ \psi^{\dagger}_{-} B_{+}^a t^a \psi _{-} + \psi^{\dagger}_{-} C_{+}
\psi _{-}] \}  \nonumber \\
&=& \int \! D B^a_{+} D \lambda_{-}^a DC_+ D\rho_- \exp \{i\int \! d^2 x \,[
\psi^{\dagger}_{-} B_{+}^a t^a \psi _{-} +B_{+}^a \lambda_{-}^a
+\psi^{\dagger}_{-} C_{+} \psi _{-} +C_{+} \rho_{-}] \}\quad.
\end{eqnarray}
}

{\small The fields $C_+$ and $\rho_-$ are introduced to factor out the $U(1)$
symmetry of the fermions, if this is neglected, a $U(N)/SU(N)$ coset model
is factored out. From now on we use the notation $B_+=B_+^at^a$ and $%
\lambda_-=\lambda_-^at^a$.  The procedure is formally the same as in the
abelian case. We shift $B_+ \rightarrow B_+ + A_+$ and perform the
non-abelian analogue of the change of variables (\ref{4}) %
\begin{equation}  \label{27}
\begin{array}{rclcrcl}
\psi_{-} & \rightarrow & e^{-i\theta}g^{-1} \psi_{-} & \quad \quad & \psi_{+}
& \rightarrow & \psi_{+} \nonumber \\ 
\psi^{\dagger }_{-} & \rightarrow & \psi^{\dagger}_{-} g e^{i\theta} & 
\quad\quad & \psi^{\dagger } _{+} & \rightarrow & \psi^{\dagger}_{+}
\end{array}
\quad,
\end{equation}
\begin{equation}  \label{28}
\lambda_-=\frac{1}{4 \pi}g^{-1}\partial_-g
\end{equation}
\begin{equation}  \label{29}
\rho_- =\frac{1}{ \pi}\partial_-\theta\quad.
\end{equation}
The contribution to the measure is the non-abelian anomaly which may be
iterated to yield \cite{physrev}, 
\begin{equation}  \label{30}
J_{{\rm F}} = \exp\{ iS[g]+\frac{i}{4\pi}\int\!d^2 x\,[{\rm Tr}(
B_{+}(\partial_{-} g)g^{-1}) +\frac{N}{2 \pi}\partial_-\theta\partial_+
\theta + \frac{N}{\pi}C_+\partial_-\theta] \}\quad,
\end{equation}
where $S[g]$ is the WZW action 
\begin{equation}  \label{31}
S[g] = - \frac{1}{8\pi}\int\!d^2 x\,{\rm Tr}(g\partial_{\mu}
g^{-1}g\partial^{\mu} g^{-1}) -\frac{1}{12\pi}\int_{\Gamma}\!d^3 x\,
\epsilon_{\mu \nu \rho} {\rm Tr}(g\partial^{\mu} g^{-1}g\partial^{\nu}g^{-1}
g\partial^{\rho} g^{-1})\quad.
\end{equation}
}

{\small Eq. (\ref{30}) is the non-abelian analogue of (\ref{6}). The
Jacobian associated with transformation (\ref{28}) is (see refs. \cite{10,13}%
) 
\begin{equation}  \label{32}
J=\int D b_+Dc_+e^{i\int \! d^2x {\rm Tr}(b_+\partial_-c_+)} \quad,
\end{equation}
where Grassmann fields have been introduced in the adjoint of $SU(N)$. For
the change of variables (\ref{29}) we have the Jacobian (\ref{7}). }

{\small Transformations (\ref{28}) and (\ref{29}) were chosen so that the $%
B_+^a \lambda_-^a$ term of eq. (\ref{24}) cancels against the $%
B_+g^{-1}\partial_- g $ term of eq. (\ref{30}). Shifting the $B_+$ field to $%
B_+ \rightarrow B_+ - ig^{-1} \partial g$ and $C_+ \rightarrow C_+ -
\partial_+\theta $, we obtain a completely decoupled fermionic coset model
with $B_+$ constraining the $SU(N)$ currents to be zero and $C$ constraining
the $U(1)$ current to be zero. The dynamics is contained in the WZW and free
boson model, 
\begin{equation}  \label{34}
Z[A_+,{\bar \eta},\eta]_{\bar{\eta},\eta =0} =Z_{{\rm coset}}Z_{{\rm B}}[A_+,%
{\bar \eta},\eta]_{\bar{\eta},\eta =0}
\end{equation}
%
%
where 
\begin{equation}  \label{35}
Z_{{\rm coset}} = \int \! D \bar{\psi} D \psi D B_{+} DC_+ D b_+ D c_+ D {%
\tilde b}_+ D {\tilde c}_+ \, e^{iS_{{\rm coset}}}\, ,
\end{equation}
\begin{eqnarray}
S_{{\rm coset}} & = & \int \! d^2 x [ {\bar \psi}\gamma^{\mu}i\partial_{\mu}
\psi + \psi_{-}^{\dagger}B_{+}\psi_{-} + b _+\partial_{-} c_+  \nonumber \\
&+&\psi_{-}^{\dag}C_{+}\psi_{-} + {\tilde b} _+\partial_{-} {\tilde c}_+ ]\,
,
\end{eqnarray}
and 
\begin{equation}  \label{36}
Z_{{\rm B}}[A_+,{\bar \eta},\eta] = \int Dg D\theta e^{i S_{{\rm B}}}
\end{equation}
with 
\begin{eqnarray}
S_{{\rm B}}&=&WZW[g]+ \int\!d^2x [ \frac{1}{4\pi}{\rm Tr}(
A_{+}(\partial_{-} g)g^{-1}) +\frac{N}{2\pi}\partial_-\theta\partial_+\theta
\nonumber \\
&+&\eta\psi^{\dagger j}_+g_{jk}e^{i\theta}\psi_-^{k}+{\bar \eta}%
\psi^{\dagger j}_-g^{-1}_{jk} e^{-i\theta}\psi_+^{k} ] \quad.
\end{eqnarray}
}

{\small The only coupling between the coset model and the boson model is
through the sources of the mass bilinears. Functional differentiation with
respect to these sources at zero gives a complete decoupling between the
coset and boson models, as indicated in expression (\ref{34}). As before,
the fermionic correlators are constants (because the coset model has $c=0$ 
\cite{12}),  which corresponds on the second quantized level to fermionic
operators independent of space-time. The Hilbert space of a $SU(N)$ coset
model is finite dimensional, and the $U(1)$ coset model has a discrete
Hilbert space. We thus retain this fermionic Hilbert space in our
bosonization prescription. The role of this space is to ensure the correct
superselection rules, while the dynamics is in the bosonic sector. }

{\small If we have both colour and flavour indices on the fermions we may
proceed in two different ways. Consider the Lagrangian %
%
%
\begin{equation}  \label{37}
L = {\bar \psi}^j_{\alpha}\gamma^{\mu}i\partial_{\mu} \psi^j_{\alpha} + \eta
\psi_{+\; \alpha}^{\dagger j} \psi_{- \;\alpha}^j + {\bar \eta} \psi_{-
\;\alpha}^{\dagger j} \psi_{+ \;\alpha }^j \quad,
\end{equation}
where $j=1 \ldots N_f$ is called a flavour index and $\alpha = 1 \ldots N_c $
a colour index. We focus our attention here on source terms for fermionic
masses as the bosonization rule for the relevant currents is
straightforward. The model actually has an $U(N_f N_c)$ symmetry which is
larger than the $U(1)\times SU(N_f)\times SU(N_c)$ symmetry where we limit
ourselves to separate rotations of the flavour and colour indices. The $%
U(N_f N_c)$ bosonization scheme of \cite{5,6} is reproduced by factoring out
an $U(1)\times SU(N_f N_c)$ coset model. This is done precisely as described
above with $g$ taking values in $SU(N_f N_c)$ as may be seen by relabling
the fermions with a single index $r=1 \ldots N_f N_c$. }

{\small Alternatively we may factor out an $U(1)\times SU(N_f)\times SU(N_c)$
fermionic coset by generalizing the identity (\ref{26}) to 
\begin{eqnarray}  \label{38}
1& =& \int \! D B_{+}^a \delta( B_{+}^a) DE^b_+\delta(E^b_+)DC_+\delta(C_+) 
\nonumber \\
& &\exp\{i\int \! d^2 x \, [ \psi^{\dagger}_{-} B_{+}^a T^a \psi _{-}
+\psi^{\dagger}_{-} E_{+}^b t^b \psi _{-} + \psi^{\dagger}_{-} C_{+} \psi
_{-}] \}
\end{eqnarray}
where $T^a$ are generators of $SU(N_f)$ and $t^a$ are generators of $SU(N_c)$%
. The Dirac deltas are handled, as usual, by introducing Lagrange
multipliers, and the chiral transformation in the present case reads 
\begin{equation}  \label{39}
\begin{array}{rclcrcl}
\psi_{-} & \rightarrow & e^{-i\theta}g^{-1}h^{-1} \psi_{-} & \quad \quad & 
\psi_{+} & \rightarrow & \psi_{+} \nonumber \\ 
\psi^{\dagger }_{-} & \rightarrow & \psi^{\dagger}_{-}h g e^{i\theta} & 
\quad\quad & \psi^{\dagger}_{+} & \rightarrow & \psi^{\dagger}_{+}
\end{array}
\quad \, ,
\end{equation}
with $g \in SU(N_f)$ and $h \in SU(N_c)$. By performing again the relevant
changes of variables on the Lagrange multipliers (transforming from the
algebra to the group) and shifting the auxilliary fields, the final result
is easily seen to be 
\begin{equation}  \label{40}
Z[{\bar \eta},\eta]_ {\eta,\bar\eta=0} =Z_{{\rm coset}}Z_{{\rm B}}[{\bar \eta},\eta] _{\eta,\bar\eta=0} 
\end{equation}
where the coset action is given by 
\begin{equation}
S_{{\rm coset}} = \int \! d^2 x [ {\bar \psi}\gamma^{\mu}i\partial_{\mu}
\psi + \psi_{-}^{\dagger}B_{+}^a T^a\psi_{-} + \psi_{-}^{\dagger}E_{+}^b
t^b\psi_{-}+ {\rm ghosts}] \quad,
\end{equation}
and the bosonic action by 
\begin{eqnarray}  \label{41}
S_{{\rm B}}&=& N_c\;WZW[g]+N_f\;WZW[h] +\int\!d^2x[ \frac{N_cN_f}{2 \pi}%
\partial_-\theta\partial_+\theta  \nonumber \\
& +&\eta\psi^{\dagger l}_{+ \;\alpha}g_{lj}h_{\alpha
\beta}e^{i\theta}\psi_{-\;\beta} ^{j} +{\bar \eta}\psi^{\dagger l}_{-
\;\alpha}g^{-1}_{lj}h^{-1}_{\alpha\beta}e^{-i\theta}\psi_{+\;\beta}^{j} ]
\quad.
\end{eqnarray}
}

{\small The only new aspect in the derivation of (\ref{40}) is to remember
that we pick up $N_f$ times the colour anomaly, $N_c$ times the flavour
anomaly and $N_f N_c$ times the abelian anomaly under transformation (\ref
{39}). }

{\small As indicated in (\ref{40}), functional differentiation with respect to the sources
and equating them to zero give us complete decoupling. For example the two
point mass correlator reads %
\begin{eqnarray}  \label{42}
&&\langle\; \psi^{\dagger}_{+}\psi_{-}(x) \psi^{\dagger}_{-}
\psi_{+}(y)\;\rangle_{{\rm {free\,\, fermion}}} = \langle\; \psi^{\dagger
i}_{+\; \alpha}\psi_{-\;\beta}^j(x) \psi^{\dagger k}_{-\;\gamma}
\psi^l_{+\;\delta}(y)\;\rangle_{{\rm {coset}}}  \nonumber \\
&&\quad \times\langle\;g_{ij}(x)g^{-1}_{kl}(y)\;\rangle \langle \;h_{\alpha
\beta}(x)h^{-1}_{\gamma \delta}(y)\rangle \langle\;e^{ i \theta(x)}e^{- i
\theta(y}\;\rangle_{{\rm {free\,\, boson}}}\,\quad.
\end{eqnarray}
The fermionic coset part is again a trivial model with central charge $c=0$ 
\cite{12}, enforcing only superselection rules. }

{\small The result (\ref{42}) differs from the bosonic representation of the
fermionic mass bilinears given in Refs.\cite{5,6} by the presence of the
coset operator $\psi_{\alpha}^j$. These operators still carry both flavour
and colour indices and therefore we do not have the complete decoupling of
colour and flavour degrees of freedom. In Refs.\cite{5,6} it was noted that
the omission of these factors leads to an incorrect result for the four
point function (of mass bilinears) and in  Ref. \cite{5} it was suggested
that a finite dimensional Hilbert space should be retained on the bosonic
level to solve this problem. Our approach  automatically provides the coset
Hilbert space that plays this role. }

{\small The coset model  is topological as it contains only the zero modes
of the free fermions. The result (\ref{41}) therefore suggests the following
boson image for the mass bilinear on the second quantize level 
\begin{equation}  \label{43}
\psi^{\dagger p}_{+ \;\gamma} \psi_{-\;\gamma} ^{p} = ( c^{\dagger l}_{+
\;\alpha}+ d^{ l}_{+ \;\alpha} )(c_{-\;\beta} ^{j} + d_{-\;\beta} ^{\dagger
j}) g_{lj}h_{\alpha \beta}e^{i\theta} \quad,
\end{equation}
where $c$ and $d$ are fermion creation and annihilation operators 
\begin{eqnarray}
\{ c_{a\;\alpha} ^{j}, c^{\dagger l}_{b \;\beta} \}& =&\delta^{j l}\delta_{a
b}\delta_{\alpha \beta}  \nonumber \\
\{ d_{a\;\alpha} ^{j}, d^{\dagger l} \}& =&\delta^{j l}\delta_{a
b}\delta_{\alpha \beta}
\end{eqnarray}
with $a$ and $b$ a left or right index. Other anticommutators vanish, and
the fermion Fock vacuum is defined by $c|0\rangle=d|0\rangle=0$. }

{\small In the case of the $U(N_f N_c)$ bosonization scheme (\ref{35}) the
fermionic coset operators should, in principal, also be retained in the
bosonization dictionary. However, in this case the $SU(N_f N_c)$ selection
rules are simpler and little harm is done ignoring this factor, as the $%
SU(N_f N_c)$ selection rules are also present for the WZW model. }

\section{\protect\small Conclusions}

{\small In this paper we have shown how various bosonization schemes that
exist in two dimensional quantum field theory may be understood from the
path integral point of view in terms of a single formalism where a trivial
fermionic coset model is factored out. We showed how the old non-abelian
bosonization may be incorporated in the path integral approach and applied
this to the massless $SU(2)$ Thirring model to show its equivalence to the
sine-Gordon model and a free boson model. }

{\small It is known from the old bosonization dictionary that constant
fermionic operators are required on the bosonic level. The only way of
incorporating this notion of constant operators in a path integral is in
terms of a field theoretic model that is dynamically trivial or topological.
We are naturally lead in our approach to fermionic coset models that play
this role. }

{\small This path integral approach to bosonization forces us to keep a
finite dimensional fermionic Hilbert space (apart from the $U(1)$ which is
discrete), which enforces superselection rules while the dynamics decouples
into the bosonic part. The fermionic coset Hilbert space may be neglected
when the selection rules are obvious, but in the general case it must be
retained. Two particular cases where this is important is in the old
non-abelian bosonization (abelian bosonization applied to non-abelian
models) and in the $SU(N_f)\times SU(N_c)$ non-abelian bosonization scheme
of coloured and flavoured fermions. }

{\small {\bf Acknowledgements} }

{\small The authors acknowledge support from the Foundation for Research
Development. }

\end{document}